\documentclass[onecolumn]{aastex63}

\usepackage{lineno}
\usepackage{epstopdf}
\usepackage{appendix}
\usepackage{color}
\usepackage{url}
\usepackage{mathrsfs}
\usepackage{indentfirst}
\usepackage{latexsym,bm}
\usepackage{amsmath}

\usepackage{hyperref}

\shorttitle{A 31.3 day QPO in the blazar S5 0716+714}
\shortauthors{Chen et al.}

\begin{document}

\title{A 31.3 day Transient Quasiperiodic Oscillation in Gamma-ray Emission from Blazar S5 0716+714}

\correspondingauthor{Tingfeng Yi}
\email{ytf@ynnu.edu.cn}

\author{Junping Chen}
\affiliation{Key Laboratory of Colleges and Universities in Yunnan Province for High-energy Astrophysics, Department of Physics, Yunnan Normal University, Kunming 650500, People's Republic of China}
\affiliation{Guangxi Key Laboratory for the Relativistic Astrophysics, Nanning 530004, People's Republic of China}

\author[0000-0001-8920-0073]{Tingfeng Yi}
\affiliation{Key Laboratory of Colleges and Universities in Yunnan Province for High-energy Astrophysics, Department of Physics, Yunnan Normal University, Kunming 650500, People's Republic of China}
\affiliation{Guangxi Key Laboratory for the Relativistic Astrophysics, Nanning 530004, People's Republic of China}

\author{Yunlu Gong}
\affiliation{Department of Astronomy, School of Physics and Astronomy, Key Laboratory of Astroparticle Physics of Yunnan Province, Yunnan University, Kunming 650504, People's Republic of China}

\author{Xing Yang}
\affiliation{Guangxi Key Laboratory for the Relativistic Astrophysics, Nanning 530004, People's Republic of China}
\affiliation{School of Physical Science and Technology,Guangxi University, Nanning 530004, People's Republic of China}

\author{Zhihui Chen}
\affiliation{Key Laboratory of Colleges and Universities in Yunnan Province for High-energy Astrophysics, Department of Physics, Yunnan Normal University, Kunming 650500, People's Republic of China}


\author{Xin Chang}
\affiliation{Key Laboratory of Colleges and Universities in Yunnan Province for High-energy Astrophysics, Department of Physics, Yunnan Normal University, Kunming 650500, People's Republic of China}

\author[0000-0002-2720-3604]{Lisheng Mao}
\affiliation{Key Laboratory of Colleges and Universities in Yunnan Province for High-energy Astrophysics, Department of Physics, Yunnan Normal University, Kunming 650500, People's Republic of China}

\begin{abstract}
We systematically search for quasiperiodic oscillatory (QPO) signals on the month timescale among the 1525 sources given in the Fermi Large Area Telescope Light Curve Repository. We find a transient QPO of 31.3$\pm$1.8 days in the gamma-ray band light curve of the TeV blazar S5 0716+714, which has seven cycles (MJD 55918-56137) for the first time by weighted wavelet Z-transform and Lomb-Scargle periodogram methods. Monte Carlo simulations based on the power spectral density and probability distribution function were used to evaluate the confidence level of the QPO, and the result is $\sim 4.1\sigma$. Seasonal autoregressive integrated moving average modeling of the light curve revealed it is a significant physical QPO. The physical models to explain the sporadic QPO of the month-timescale QPOs in blazar were discussed. Our studies indicate that the helical jet model and blob move helically in a curved jet model to properly explain this kind of transient QPO.
\end{abstract}
\keywords{active galactic nuclei: gamma-rays - BL Lacertae objects: individual: S5 0716+714 - quasi-periodic oscillation}
\section{Introduction}\label{sec:intro}
Active galactic nuclei (AGN) are energetic astrophysical sources that are powered by the accretion of gas onto the super-massive black hole (SMBH) at its center. It shows unique radiation characteristics, and all band radiation from radio to high-energy gamma-ray has been observed \citep{2017A&ARv..25....2P}. Blazar is a subclass of AGN, whose relativistic jet orientates the observer's line of sight. According to the intensity of the emission line in the spectrum, a blazar is further divided into BL Lacertae (BL Lac) object and flat spectrum radio quasar (FSRQ). The spectrum of BL Lac contains very weak and narrow emission lines, while FSRQs show wide and strong emission lines \citep{1995PASP..107..803U,2011MNRAS.414.2674G}.

Based on the Fermi Gamma-ray Space Telescope long-term survey data, a few of AGN's QPOs were reported in the gamma-ray band, ranging from days to months and even years of timescales.
For the year-like QPOs, the physical mechanism of them was mostly attributed to pulsating accretion flow instability produces precession or helical jets, a homogeneous curved helical jet scenario, Lense-Thirring precession of the flow, Keplerian binary orbital motion would induce periodic accretion perturbations \citep{2015ApJ...813L..41A}. Several cases were reported, such as, PG 1553+113, PKS 0426-380, PKS 0301-243, PKS 2155-304, PKS 0537-441, PMN J0948+0022, OJ 287, PKS 0601-70, PKS 0521-36, 4FGL J0112.1+2245 \citep{2015ApJ...813L..41A,2017ApJ...835..260Z,2017ApJ...845...82Z,2017ApJ...842...10Z,2016ApJ...820...20S,2017ApJ...849...42Z,2020MNRAS.499..653K,2020ApJ...891..163Z,2021ApJ...919...58Z,2022Ap&SS.367....6G}.
For the month-like QPOs, the physical mechanism for the generation of these QPOs is mainly attributed to the helical structure in the jet \citep{2018NatCo...9.4599Z}. For example, PKS 2247-131, B2 1520+31, PKS 1510-089, and 3C 454.3 were reported to have 34.5, 71, 92 and 47 days QPO respectively \citep{2018NatCo...9.4599Z,2019MNRAS.484.5785G,2022MNRAS.510.3641R,2021MNRAS.501...50S}.
For more short (the day-like) QPOs, the physical mechanism of them is flux enhancements by the magnetic reconnection events in the magnetic islands inside the jet. This kind of QPO was reported in two-source, i.e., CTA 102 ($\sim 7.6$ days) and PKS 1510-089 ($\sim 3.6$ days) \citep{2020A&A...642A.129S,2022MNRAS.510.3641R}.

The QPO signals of AGN are also found in optical, X-rays, and radio bands. The most well-known case is BL Lac OJ 287, which exhibits $\sim$12 years of QPO in its optical light curve based on more than a century of monitoring \citep{1992A&A...264...32K,2006ApJ...646...36V}. In addition, AO 0235+16, AO 0235+164, S5 0716+714, and PKS 2155-304 are reported to have QPO signals in the optical band.
It was found that these QPO signals are distributed on the time scales of the hours, days, and years \citep{2001A&A...377..396R,2006ApJ...650..749L,2009ApJ...690..216G,2009A&A...506L..17L,2014RAA....14..933Z}.
RE J1034+396, 2XMM J123103.2+110648, 1H 0707-495, and Mrk 766 are reported to have QPO signals in X-rays. Unlike those found in gamma-ray, these X-ray QPO signals oscillate on time scales of hours or days, and no reliable QPO signals have been reported on long time scales (year-like oscillation) \citep{2008Natur.455..369G,2013ApJ...776L..10L,2016ApJ...819L..19P,2017ApJ...849....9Z,2018ApJ...853..193Z,1998ApJ...492L..59S}. FSRQ J1359+4011, PKS J0805-0111 and BL Lac PKS 0219-164, PKS J2134-0153 are reported to have QPO signals in the radio band, most of which are on the time scale of years \citep{2013MNRAS.436L.114K,2017ApJ...847....7B,2021MNRAS.506.3791R,2021RAA....21...75R}.

S5 0716$+$714 is a very famous BL Lac object, which has been monitored by many telescopes for a long time, and its redshift z = 0.31 \citep{2001A&A...375..739D}. S5 0716$+$714 is found to have intra-day periodic oscillation, and the mass of the central black hole is estimated to be $>2.5\times10^6\rm{M_{\bigodot}}$, no correlation between the flux level and the time scale of intra-day variability (IDV) was found, so it is speculated that there is more than one radiation mechanism in this source \citep{2009ApJ...690..216G}. There is a significant correlation between the change of gamma-ray flux and the change of position angle in the jet measured by very long baseline interferometry \citep{2014A&A...571L...2R}. From January 19 to February 22, 2015, swift/XRT monitored the continuous burst status of S5 0716$+$714. The observed X-ray spectra conform to the breaking power-law model. With the increase of flux, the braking energy is transferred to higher energy, which indicates that synchrotron radiation is dominant in the observed X-ray spectra \citep{2015MNRAS.452L..11W}. NuStar observed the hard X-ray of S5 0716$+$714 and found a clear energy band fracture (8 keV) \citep{2016MNRAS.458.2350W}. S5 0716$+$714 has obvious optical IDV, and multiple intra-day QPO evidence has been found in multiple monitoring in different time epochs \citep{2017MNRAS.469.2457L,2017A&A...605A..43Y,2018AJ....155...31H,2018AJ....156...36K,2019ApJ...880..155L,2021RAA....21..102L}. In the burst stage of S5 0716$+$714, the gamma-ray spectrum shows an obvious broken feature between 0.93 and 6.90 GeV, and five photons with energy $E>0.1$ TeV are found \citep{2020ApJ...904...67G}. The TeV emission from S5 0716+714 originated in a superluminal knot, which means that the high-energy radiation may come from the moving emission region in the helical path upstream of the jet \citep{2018A&A...619A..45M}. The space telescope (Transiting Exoplanet Survey Satellite, TESS) in 2019 December-2020 January observed and analyzed the fast variability of the object with unprecedented high-resolution sampling, revealing the characteristics of IDV \citep{2021MNRAS.501.1100R}. Recent studies have shown that there is a significantly positive or negative correlation between S5 0716+714 radio bands (15, 37, and 230 GHz) and gamma-ray (0.1-200 GeV) fluxes in different time ranges \citep{2022ApJ...925...64K}.

Among 1525 sources with a variation index greater than 21.67 given in the Fermi Large Area Telescope Light Curve Repository (\href{https://fermi.gsfc.nasa.gov/ssc/data/access/lat/LightCurveRepository/}{LCR}), we search for the QPO on a time scale of months. The results show that S5 0716+714 is another blazar with months QPO, in addition to the previously reported PKS 2247-131, B2 1520+31, PKS 1510-089, and 3C 454.3. S5 0716+714 has a $\sim31.3$ days QPO with a $4.1\sigma$ confidence level, based on a Large Area Telescope (Fermi-LAT) light curve [Modified Julian Day (MJD): 55918-56137] with 5 days bin. Fermi-LAT data reduction and results of searching for periodicity by different methods are presented in section 2. The conclusion and discussion are given in section 3.

\section{DATA ANALYSIS AND RESULTS}
\subsection{Fermi-LAT Data}

The \href{https://fermi.gsfc.nasa.gov/ssc/data/access/}{Fermi-LAT}, the primary instrument of the Fermi satellite mission, is an imaging, wide field-of-view (FoV), high-energy gamma-ray telescope, covering the energy range from below 20 MeV to more than 300 GeV. LAT has a large angular field of view of about 2.3 sr and covers the entire sky every three hours and a large effective area of $>$ 8000 $cm^2$ at $\sim$1 GeV \citep{2009ApJ...697.1071A}.

For data selection, we selected LAT events in the Mission Elapsed Time (MET) 239557417.0-647393497.0 ($\sim12.9$ years) time from the fermi PASS 8 database and used the Science Tools package of version $v11r05p3$ provided by Fermi Science Support Center2 (\href{https://fermi.gsfc.nasa.gov/ssc/data/analysis/software/}{FSSC}).
Firstly, we chose the events belonging to the SOURCE class (evclass=128, evtype=3) within the energy range of 0.1-300 GeV from a circular region of interest (ROI) having a radius of 15$^{\circ}$ centered at the source 4FGL J0721.9+7120 (S5 0716+714).
We excluded events with zenith angles over $90 ^{\circ}$ to improve the point-spread function and reduce diffuse emissions, and use the standard filter (DATA\_QUAL$>$0) \&\& (LAT\_CONFIG$==$1) to select a good time interval (GTI) to obtain high-quality data.
Afterward, we create the source model XML file containing the spectral shapes of all sources within the radius of the ROI+10$^{\circ}$ around the source location, including two background templates for Galactic and Extragalactic Diffuse Emissions, modeled using files $gll\_iem\_v07.fit$ and $iso\_P8R3\_SOURCE\_V3\_v1.txt$, respectively. Note that the normalization of the two diffuse emission components is set as a free parameter in the analysis.
Finally, we performed a maximum likelihood analysis of the input XML spectrum file using the $gtlike$ tool, using the instrument response function (IRF) $P8R3\_SOURCE\_V3$ to obtain the spectrum of the source.
In the 4FGL catalog, the final source spectrum is modeled using a logarithmic parabola \citep{2020ApJS..247...33A} as follows:

\begin{equation}
\frac{dN}{dE}=K(\frac{E}{E_b})^{- \alpha -\beta log(E/E_b)},
\label{eq:LebsequeIp1}
\end{equation}
where $\alpha$ is the spectral index at the break energy ($E_b$), we kept $E_b$ fixed during likelihood fitting.
The optimizer finds the most suitable spectral parameters but not the location. In other words, the fit tool does not fit the source coordinates. Therefore, Test Statistic (TS) is added here for each location to calculate the saliency of the point source at that location, $TS = -2 (lnL0 - lnL)$, where L0 represents the maximum-likelihood value fitted by the background model, and $L$ represents the maximum-likelihood value fitted by adding a test point source to the background model.

\begin{figure*}
\begin{minipage}[t]{1\textwidth}
\centering
\includegraphics[height=5cm,width=16cm]{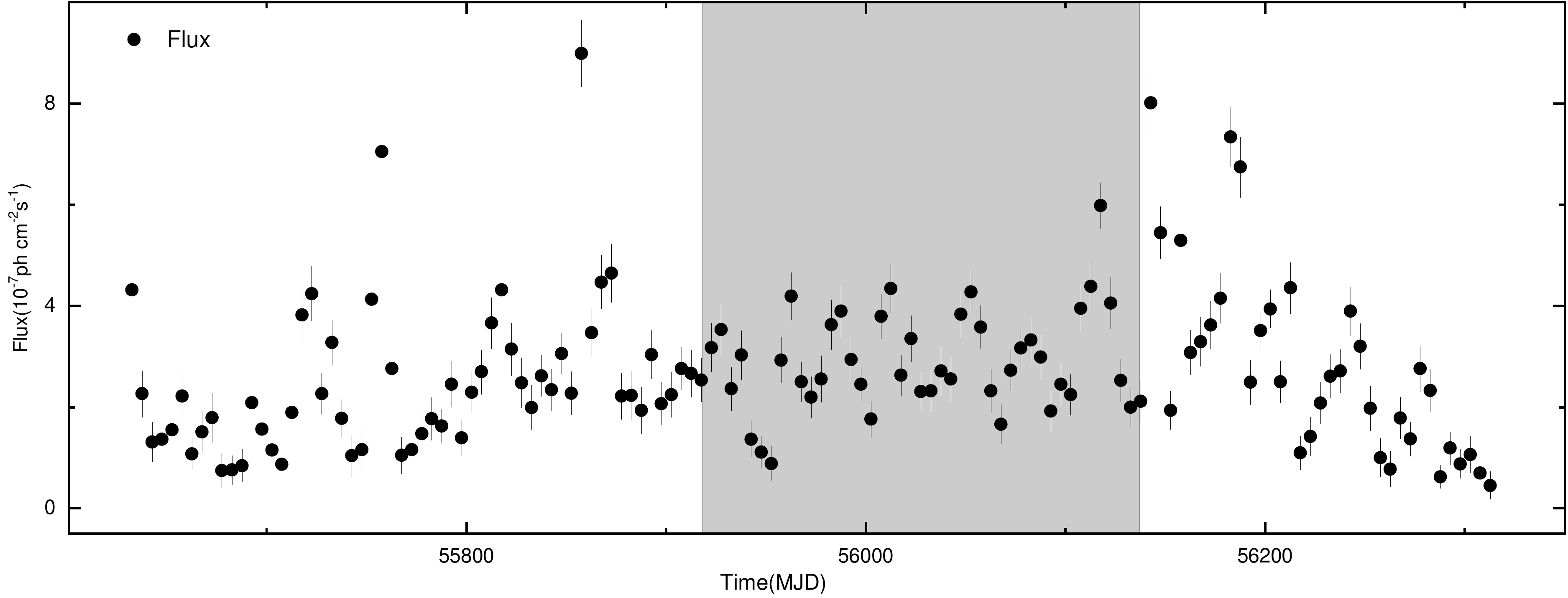}
\includegraphics[height=5.5cm,width=16cm]{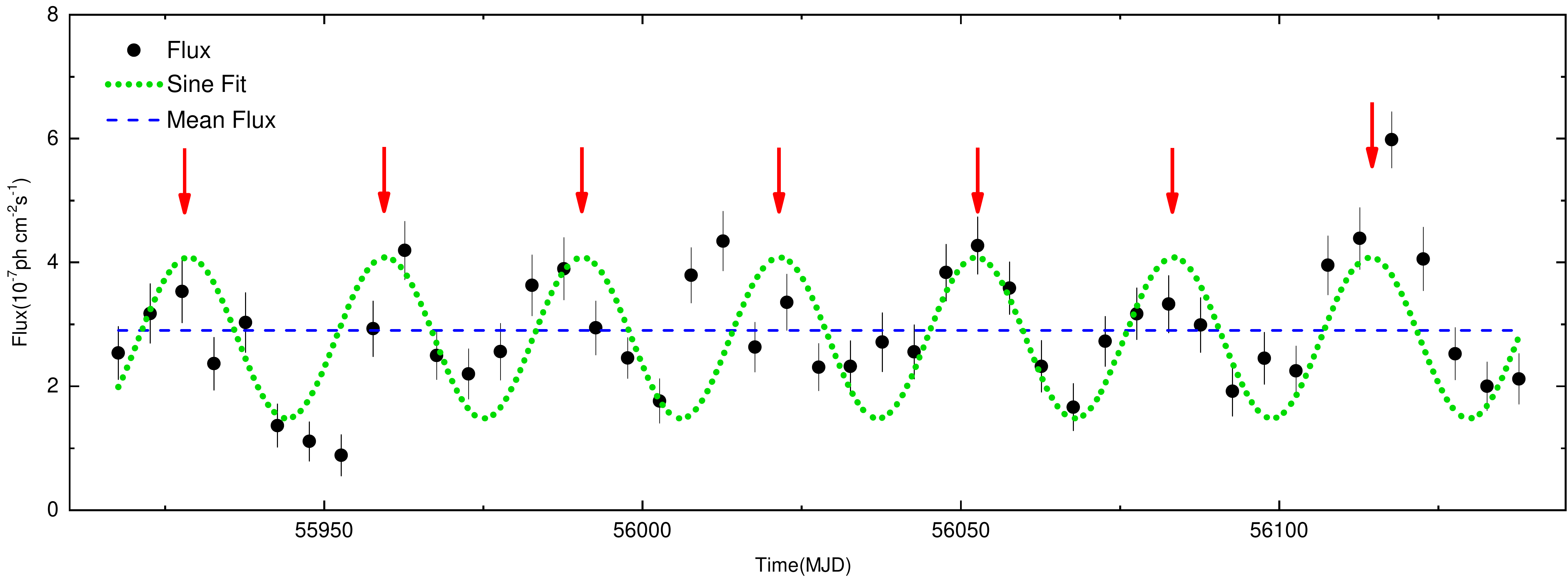}
\caption{Top panel: The light curve of Fermi-LAT, with 5 days time bin ($TS>25$) of S5 0716+714 in MJD 55593-56312. Bottom panel: This panel is the enlarged shade section (MJD 57693-57903) of the top panel. The blue dashed line represents the average value of flux; the green dotted line is the result of fitting the sine function; and the red arrows indicate the peak positions for each cycle.}
\label{Figure2}
\end{minipage}
\end{figure*}

Based on the data analysis, we generated flux data for 5 days, excluding very few data points with $TS$ less than 25. The light curve of the flaring state (MJD 55632-56312) is shown in Figure \ref{Figure2} (Top panel), and the gray shaded part (MJD 55918-56137) is shown in Figure \ref{Figure2} (Bottom panel), where the blue dotted line is the average value of the flux, and the red arrows indicate the peak positions for each cycle.
\subsection{QPO analysis and results}
\begin{figure}[h]
\begin{minipage}[t]{0.2\linewidth}
\centering
\includegraphics[height=7.1cm,width=8cm]{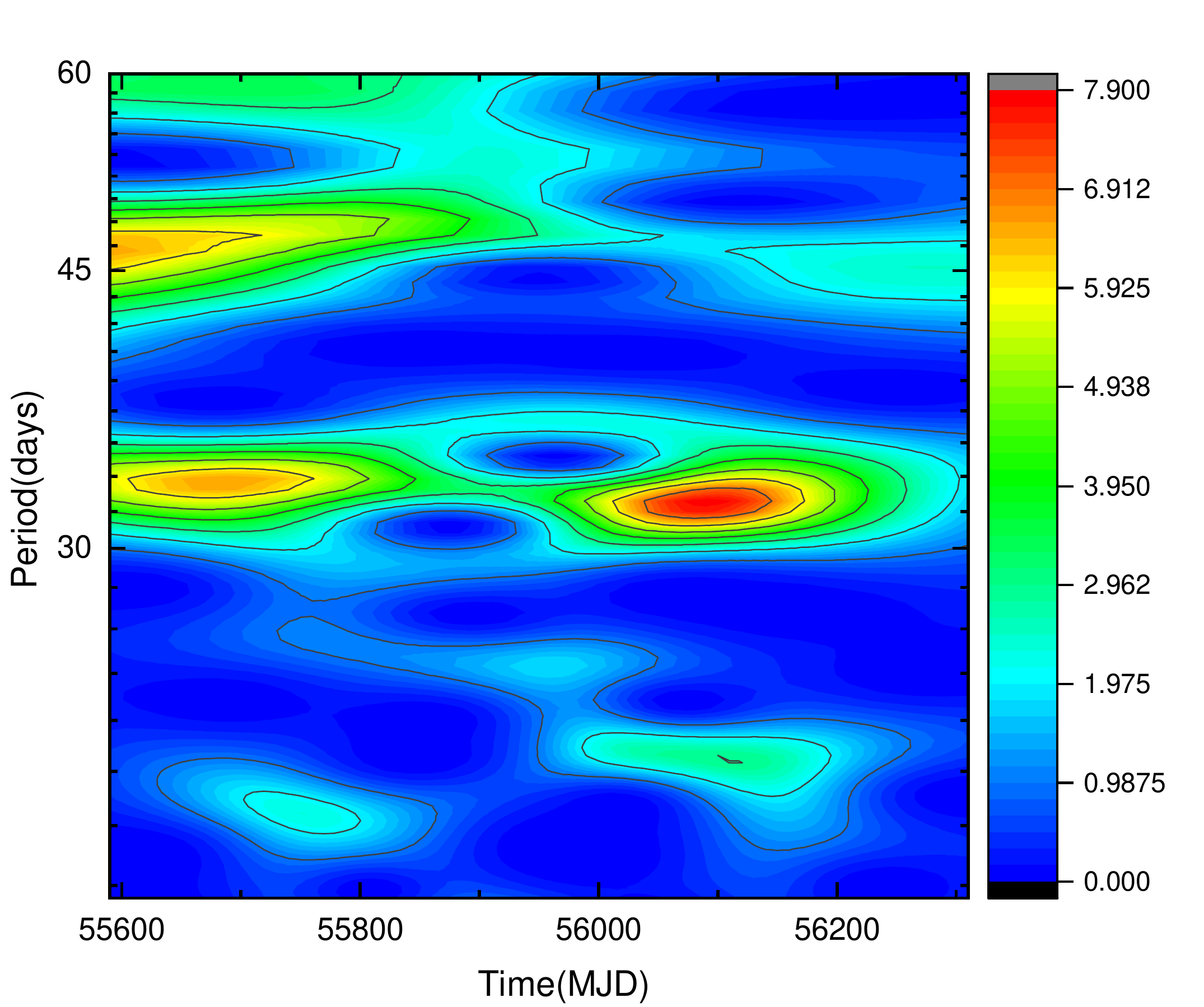}
\end{minipage}
\begin{minipage}[t]{1\linewidth}
\centering
\includegraphics[height=6.7cm,width=10cm]{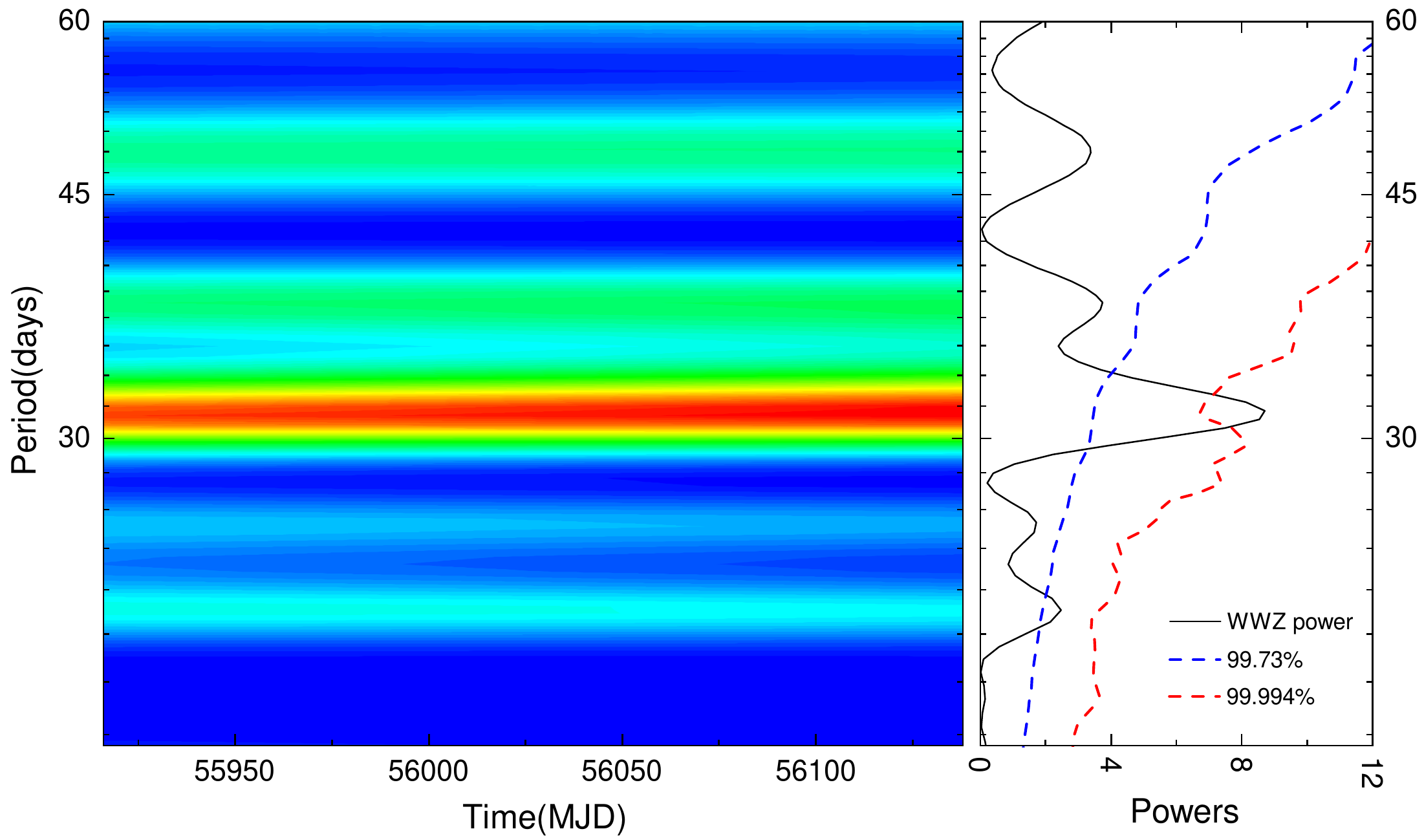}
\end{minipage}
\caption{Left panel: The WWZ map of the S5 0716+714 light curve in MJD 55593-56312. Middle panel: The enlarged map (MJD 55918-56137) of the left panel. Right panel: The black solid line shows the time-averaged WWZ. The blue and green dash lines are 3$\sigma$ and 4$\sigma$ significance curves, respectively. The QPO signal of $31.3\pm1.8$ days is detected in the period with a confidence level of 4.1$\sigma$.}
\label{Figure3}
\end{figure}

\begin{figure*}
\begin{minipage}[t]{1\textwidth}
\centering
\includegraphics[height=8cm,width=11cm]{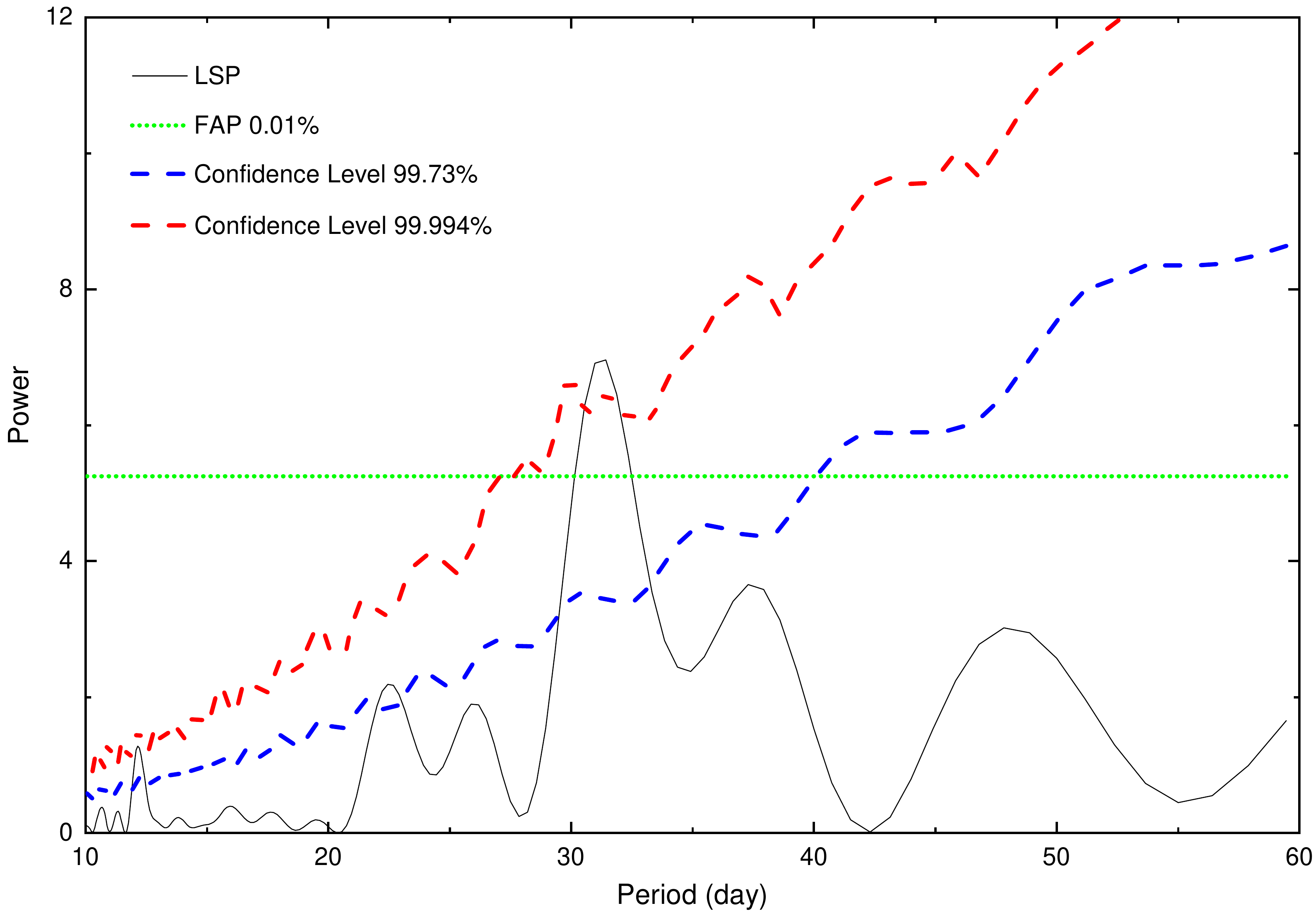}
\end{minipage}
\caption{The LSP power (black solid line) of light curve during MJD 55918-56137. The blue dash, red dash, and dotted line indicates 3$\sigma$, 4$\sigma$ and FAP (4$\sigma$), respectively.}
\label{Figure4}
\end{figure*}

\begin{figure*}
\begin{minipage}[t]{1\textwidth}
\centering
\includegraphics[height=8cm,width=11cm]{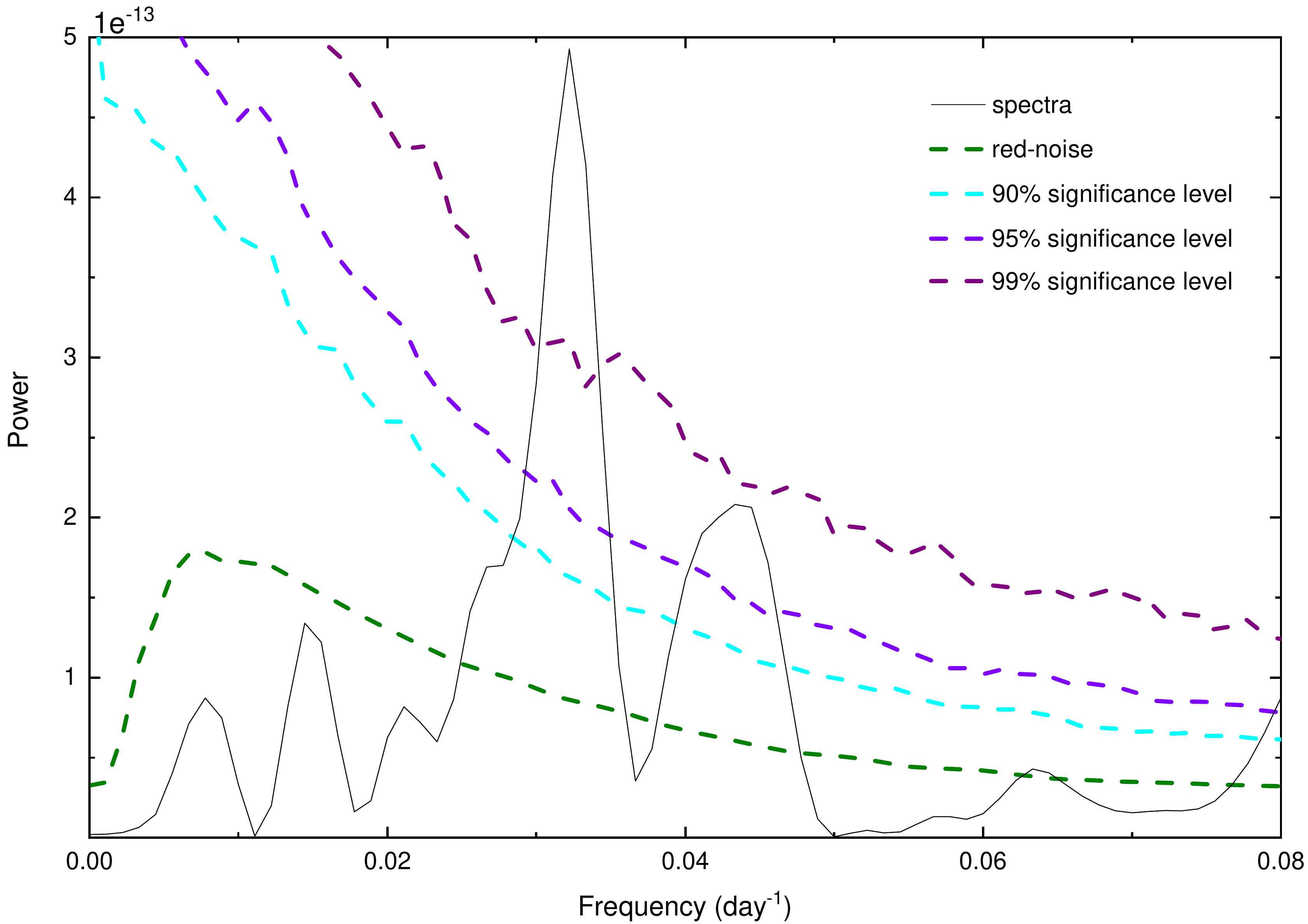}
\end{minipage}
\caption{Results of the periodicity analysis by REDFIT. The black solid line is a bias-corrected power spectrum. Dashed curves starting from the bottom are the theoretical red-noise spectrum (Olive green dash line), 90\%(cyan dash line), 95\%(violet dash line) and 99\%(purple dash line) significance levels, respectively.}
\label{Figure5}
\end{figure*}

We found a sine-like periodic variability by visual inspection, so we fitted it with a simple sine function, $y=y_0+A\sin(2\pi(t-t_c)/T)$, where $T$ is the period of the sine fitting. The fitting results show that the flux changes tend to be sinusoidal (Figure \ref{Figure2} Bottom panel, green dotted line). To study the periodicity of the gamma-ray light curve of S5 0716+714, we used three methods to analyze the QPO signal. The first method is the weighted wavelet Z-transform (WWZ)\citep{1996AJ....112.1709F}. The WWZ method can decompose the data into the time domain and frequency domain, and convolute the light curve with the kernel related to time and frequency. We used the abbreviated Morlet kernel \citep{2021MNRAS.501...50S}, and its functional form is as follows:

\begin{equation}
f[\omega(t-\tau)]=exp[i\omega(t-\tau)-c\omega^2(t-\tau)^2].
\label{eq:LebsequeIp3}
\end{equation}
The WWZ map is then given by,

\begin{equation}
W[\omega,\tau:x(t)]=\omega^{\frac{1}{2}}\int x(t)f^\ast[\omega(t-\tau)]dt.
\label{eq:LebsequeIp4}
\end{equation}
In formula (2) and (3), $f^\ast$ is the complex conjugate of the Morlet kernel $f$, $\omega$ is the frequency, and $\tau$ is the time-shift.
The advantage of WWZ map is that it can search the periodicity in the data by decomposing the signal into frequency time space at the same time, and detect any main periodicity and its duration time span. In other words, this technique measures the nonstationarity of the data set and indicates the time range of any possible periodic characteristics. For non continuous periodic signals (transient periodic signals), WWZ can show a decrease when the periodic signal starts to disappear. The advantage of the WWZ method is that it can detect any major periodicity and its duration span. We obtained the color scale of WWZ power spectrum of S5 0716+714 in the time interval, MJD 55593-56312 (the flare state), as shown in the left panel of Figure \ref{Figure3}. For the light curve during MJD 55918-56137, the WWZ method provides clearer results (see the middle panel of Figure \ref{Figure3}). WWZ power and its time average power strongly indicate the existence of QPO for about 31.3$\pm$1.8 days. The uncertainty is estimated as the half-width at half the maximum value of the Gaussian function fitting the power peak.
After that, we also used the Lomb-Scargle Periodogram (LSP) method, which is more common in astronomy, to detect the potential quasi-periodic signals in S5 0716+714. LSP method gives the power of flux modulation at different frequencies \citep{1976Ap&SS..39..447L,1982ApJ...263..835S,2009A&A...496..577Z}. For the uniformly sampled light curve, the square of the discrete Fourier transform modulus gives the periodogram. However, for irregular sampling, LSP method iteratively fits sinusoidal curves with different frequencies to light curves and constructs periodogram according to the goodness of fit, and can provide accurate frequency and power spectrum intensity. Figure \ref{Figure4} shows the power spectrum. It is obvious that there is a peak around 31.2$\pm$1.6 days. This also confirms the signal detection results of the WWZ method. Here, we use False Alarm Probability (FAP) to evaluate the confidence level of the peak. $FAP(P_n)=1- (1-prob(P>P_n))^M$, $FAP$ denotes the probability that at least one of the $M$ independent power values in a given frequency band of the white noise periodogram is greater than or equal to the power threshold $P_n$ \citep{1986ApJ...302..757H,2018ApJS..236...16V}. The green dotted line in Figure \ref{Figure4} shows the $FAP=0.01\%$, and the confidence level of the peak (31.2$\pm$1.6 days) is higher than 4$\sigma$.
The third method, REDFIT \citep{2002CG.....28..421S}, was also used to search for QPO signals in the light curve. For time-series data in the astronomy field, it is difficult to accurately estimate the red noise spectrum because the sampling time interval is usually not uniform. REDFIT was proposed to overcome this problem by fitting a first-order autoregressive (AR1) process directly to unevenly spaced time series, avoiding interpolation in the time domain and its inevitable bias. This procedure was also used to test the significance of the time series flux peaks against the background of red noise in the AR1 process. The emission fluxes of AGN are usually autoregressive \citep{2002CG.....28..421S,2020MNRAS.499..653K}, the current emission flux depends on its immediate previous emission flux, so we can use the AR1 process to model the emission red noise spectrum of S5 0716+714. The program \href{https://www.marum.de/Prof.-Dr.-michael-schulz/Michael-Schulz-Software.html/}{REDFIT3.8e} can estimate the spectrum using LSP and Welch-Overlapped-Segment-Averaging (WOSA), with a number of WOSA segments ($n50=2$), selecting a Welch window reduces spectral leakage. As shown in Figure \ref{Figure5}, there is a 31.1$\pm$1.9 days peak with a confidence level of $ >99\%$ (it is worth noting that the REDFIT3.8e code provides a maximum significance of 99\%). This result is consistent with that of the WWZ and LSP methods. However, the above three methods we use are all similar analysis and not completely independent methods to test the QPO, so we also need to consider the other models to test the physical QPO in this work.

\begin{figure}[h]
\begin{minipage}[t]{0.2\linewidth}
\centering
\includegraphics[height=7cm,width=8.5cm]{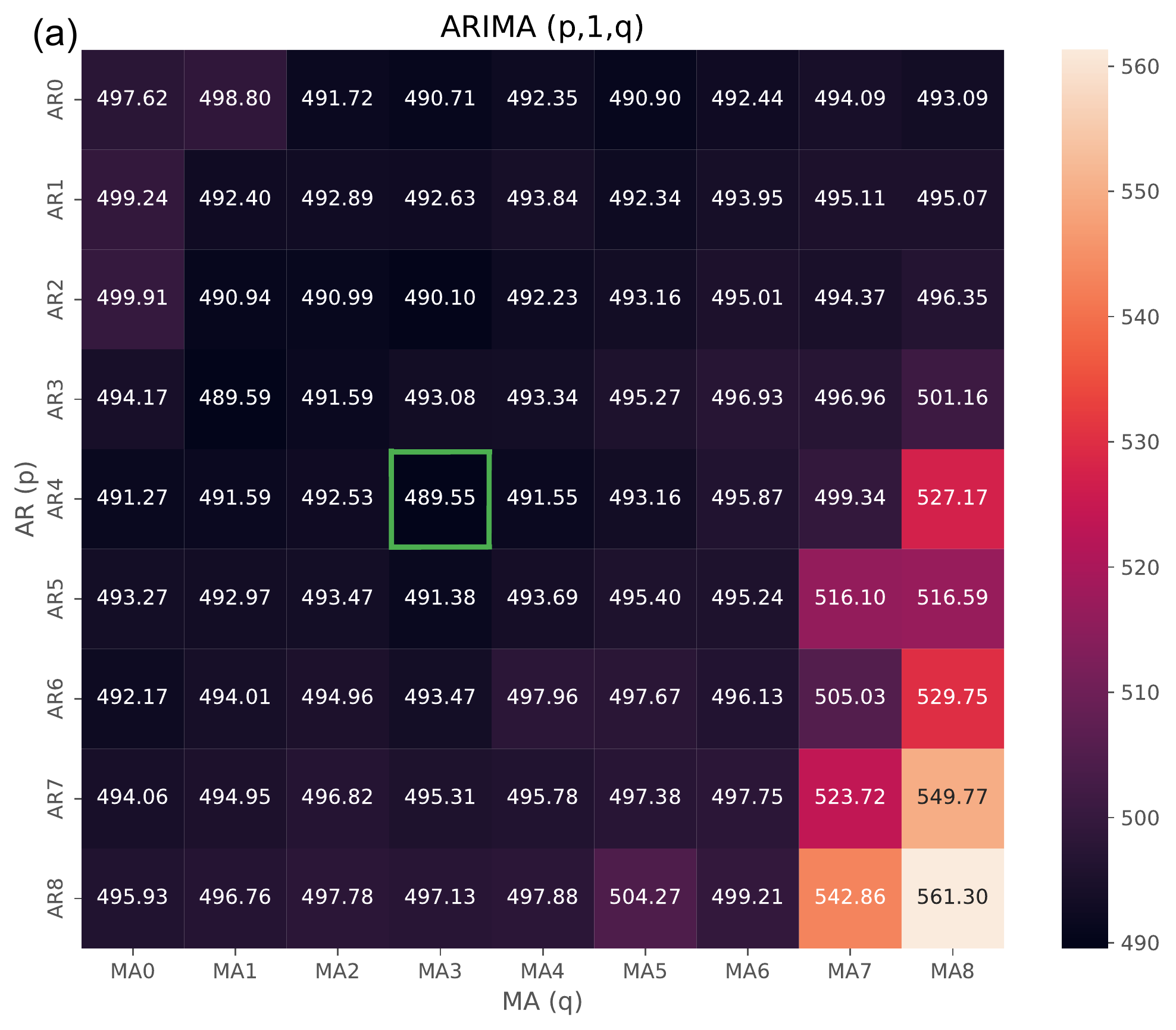}
\end{minipage}
\begin{minipage}[t]{1\linewidth}
\centering
\includegraphics[height=7cm,width=8.5cm]{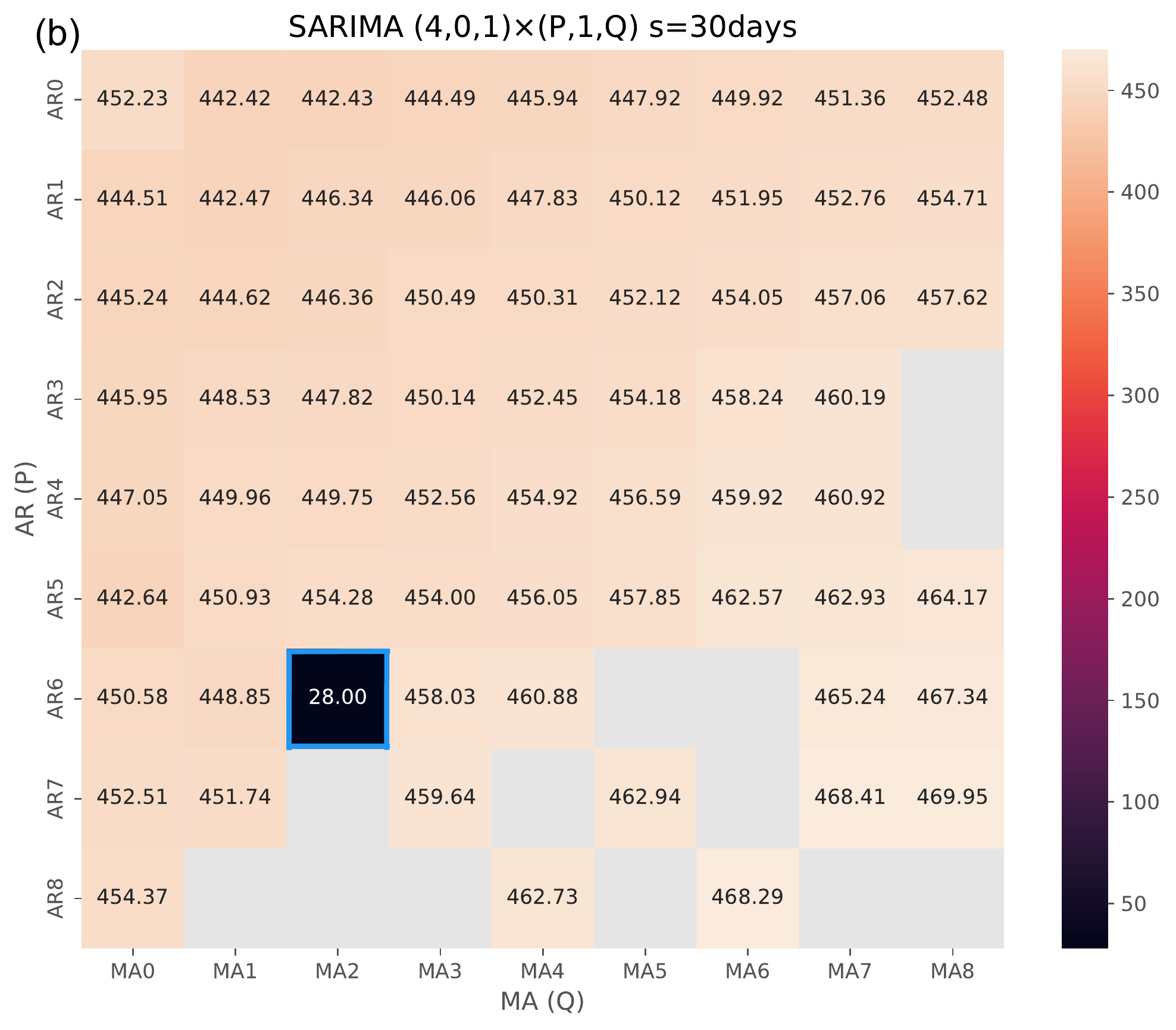}
\end{minipage}
\begin{minipage}[t]{1\linewidth}
\centering
\includegraphics[height=7cm,width=10cm]{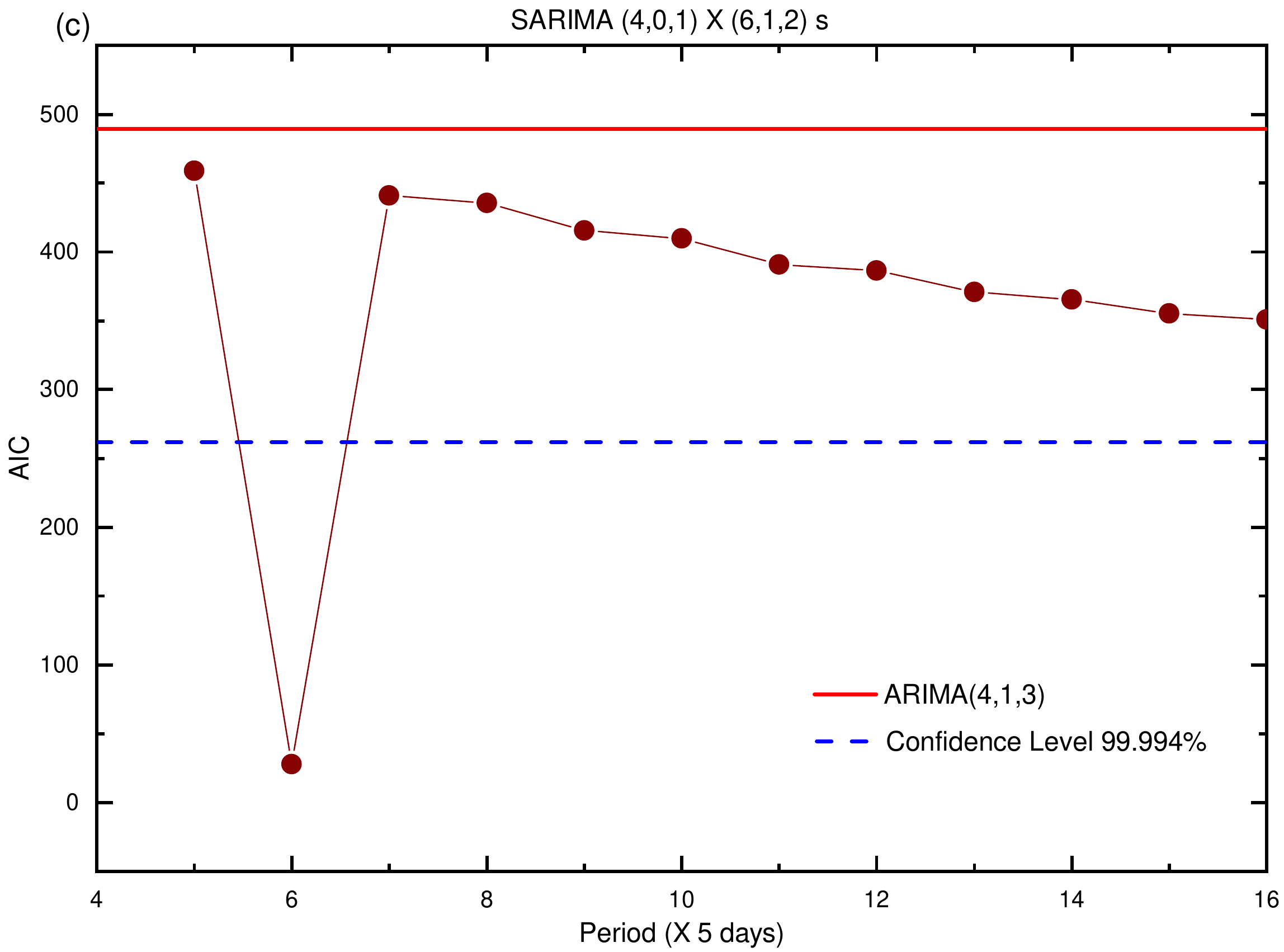}
\end{minipage}
\caption{$AIC$ distribution plot of ARIMA model and SARIMA model. (a) $AIC$ plot of the ARIMA(p, d, q) model. Although the minimum occurs at ARIMA(4, 1, 3) (the box marked with a green square), the global minimum when SARIMA is considered occurs at SARIMA(4, 0, 1)$\times$(6, 1, 2) (the box marked with blue square). (b) $AIC$ plot of SARIMA(4, 0, 1)$\times$(P, 1, Q), considering a period (30-days), we have at SARIMA(4, 0, 1)$\times$(6, 1, 2) observe a global minimum. $AIC$ values are provided in the boxes. (For boxes that are not colored and where no $AIC$ value is given, either the fit is not converging, or the $AIC$ is greater than the set threshold.) (c) $AIC$ values for different periods. We see a global minimum $AIC$ value at $s = 30$ days. The red solid line and the blue dashed line represent the ARIMA(4, 1, 3) model $AIC$ value and the 99.994\% confidence level line, respectively.}
\label{Figure6}
\end{figure}

Here, we model the light curve using the Autoregressive Integrated Moving Average (ARIMA) model, which has been widely used in disciplines or fields such as time series analysis, signal processing, and econometrics. This is a flexible method consisting of three parts, in which the autoregressive (AR) process can quantify the coefficient of the dependence of the current value on the most recent past value, the Integrated (I) is used to reduce the trend, and the moving average ( MA) process can quantify the coefficient of the current value's dependence on the system's recent random shocks \citep{1981ApJS...45....1S,1985JAWRA..21..721V,2018FrP.....6...80F}. It is defined in statistics as

\begin{equation}
\Delta^{d}F(t_{i})=\sum\limits^p_{j=1}\theta_j\Delta^{d}F(t_{i-j})+\sum\limits^q_{j=1}\phi_j\epsilon(t_{i-j})+\epsilon(t_{i}),
\label{eq:LebsequeIp6}
\end{equation}
\begin{equation}
or, \left(1-\sum\limits^p_{j=1}\theta_{j}L^{j}\right)\Delta^{d}F(t_{i})=\left(1-\sum\limits^p_{j=1}\phi_{j}L^{j}\right)\epsilon(t_{i}),
\label{eq:LebsequeIp6}
\end{equation}
where $p$, $d$, and $q$ are AR order, difference operator, and MA order respectively. The seasonal (S) autoregressive (AR) Integrated (I) moving average (MA) model or SARIMA$(p, d, q) \times (P, D, Q)_s$ is a model based on ARIMA$(p, d, q)$ evolved methods of identifying physical periodicity in light curves \citep{2013arXiv1302.6613A,2020A&A...642A.129S}, defined in statistics as

\begin{equation}
\left(1-\sum\limits^p_{j=1}\theta_{j}L^{j}\right)\left(1-\sum\limits^p_{j=1}\Theta_{j}L^{sj}\right)\Delta^{d}\Delta^{D}_{s}F(t_{i})=\left(1-\sum\limits^p_{j=1}\phi_{j}L^{j}\right)\left(1-\sum\limits^p_{j=1}\Phi_{j}L^{sj}\right)\epsilon(t_{i})+A(t),
\label{eq:LebsequeIp6}
\end{equation}
where $P$, $D$, $Q$, and $s$ are the seasonal AR order, difference operator, MA order, and period parameter, respectively. Note that the detailed derivation process for equations (4)-(6) is shown in the appendix section at the end of this paper. Next, we fit the light curves using the ARIMA and SARIMA models respectively, and compare their goodness of fit using the Akaike Information Criterion ($AIC$, \citep{1974ITAC...19..716A}). $AIC= -2lnL + 2k$, where $L$ is the likelihood function and $k$ is the number of free parameters in the model. $AIC$ rewards models that fit the data better, while penalizing models that use more parameters. Therefore, we build the parameter space to search for the model with the smallest $AIC$.

$$ \varphi=\left\{
\begin{aligned}
p,q &\in& \ [0,8] \\
d,D &\in& \ [0,1] \\
P,Q &\in& \ [0,5] \\
s   &\in& \ [0,16]\times 5 days
\end{aligned}
\right.
$$
The fitting results are shown in Figure \ref{Figure6}. When using the seasonal model SARIMA(4,0,1)$\times$(6,1,2)$_{30days}$, the $AIC$ has the global smallest value of 28.0 (Figure \ref{Figure6}b), while the $AIC$ value of the nonperiodic best-fit model ARIMA(4,1,3) is 489.55 (Figure \ref{Figure6}a), which also means that the periodic model can better fit S5 0716+714 light curve. Figure \ref{Figure6}c presents the $AIC$ values considering different periods, we find that the best $AIC$ occurs at the $30 \pm 2.5$ days position. The amplitude of the uncertainty is estimated to be half of the light curve time bin (2.5 days).
\subsection{ Significance estimation}
As described above, we used several different methods and techniques to analyze possible QPO signals in the light curve of blazar S5 0716+714 and obtained consistent results. However, the statistical characteristics of a large number of light curves in AGN show red noise-like behavior related to frequency (for example, the red noise is produced by stochastic flares), which may produce the false QPO \citep{2016MNRAS.461.3145V}. In addition, the sampling instability of the light curve (observation time, weather, seasonal variation, diurnal variation, etc.) will also produce pseudo-QPO signals. Therefore, the estimation of confidence level is very important in the QPO search. We modeled the red-noise with a first-order autoregressive process (e.g. REDFIT). The first-order autoregressive process is a pure power-law model. The pure power-laws or smooth bending power-laws could reasonably model the red-noise power spectral density (PSD), and the latter is closer to the red-noise PSD. Therefore, we further adopt a smooth bending power law to model the PSD. The Monte Carlo (MC) procedure include three-step: model the original light curve PSD and Probability Distribution Function (PDF), finally based on the parameter of PSD and PDF model to simulated light curves in the LSP and WWZ analyses \citep{2013MNRAS.433..907E}. To estimate the underlying PSD, we model the light curve red-noise PSD by smooth bending power-laws plus a Poisson noise \citep{2012A&A...544A..80G,2014MNRAS.445..437M}. Poisson noise is considered here because the observed light curve is a product of the counting detector process, so the observations are subject to Poisson noise, which is imprinted in the corresponding PSD as a constant component. The model form of PSD is

\begin{equation}
p(f:\gamma,c)=\frac{Af^{-\alpha_{low}}}{1+(f/f_{bend})^{\alpha_{high}-\alpha_{low}}} +C,
\label{eq:LebsequeIp5}
\end{equation}
where model parameters $A$, $\alpha_{low}$, $\alpha_{high}$, $f_{bend}$ and $C$ are normalization, low frequency slope, high frequency slope, bend frequency and Poisson noise, respectively. Here we obtain the optimal parameters of the PSD model through maximum likelihood analysis, i.e. firstly we calculated the joint probability density function (likelihood function) of obtaining the ensemble of periodogram estimates for the given PSD model, and then maximize the log-likelihood function probability to get the PSD model optimal parameters. For a more detailed analysis process see \citep{2013MNRAS.433..907E}. The optimal likelihood parameter values of the PSD model are $A$ ($0.014$), $\alpha_{low}$ (1.001), $\alpha_{high}$ (4.801), $f_{bend}$ ($0.056$), and $C$ ($9.638\times 10^{-5}$), respectively. The light curve in AGN generally exhibits a "burst"-like behavior, and the PDF is distributed according to a right-heavy-tailed distribution. This means that the probability of occurrence of high flux values is greater than that expected from a Gaussian distribution, so the PDF is modeled as a superposition of the two distributions of the two active states (low and flare state). This is modeled using a gamma distribution and a lognormal distribution \citep{2013MNRAS.433..907E},

\begin{equation}
PDF(x:\eta)=\omega_1\frac{x^{\alpha_s-1}e^{-x}}{\Gamma(\alpha_s)}+\omega_2\frac{1}{s(x-\mu)\sqrt{2\pi}}exp\left(-\frac{log^2(\frac{x-\mu}{\sigma})}{2s}\right),
\label{eq:LebsequeIp6}
\end{equation}
where $\alpha_s$ is the shape parameter of gamma function fitting, and the obtained value is 4.993, $log\mu$ and $\sigma$ are the log mean and standard deviation of the log-normal distribution, respectively. The maximum likelihood analysis obtains the parameter values as 2.886 and 0.401, respectively.

PSD and PDF based parameter values, we generated $10^5$ artificial light curves using the Python code \href{https://github.com/lena-lin/emmanoulopoulos}{"emmanoulopoulos"} (To account for the "red-noise leakage" effect, the generated artificial light curves are much longer (e.g. 100 times) than the original light curve dataset, where a subset with the desired length is randomly selected). Finally, according to the mean and standard deviation of the distribution of the artificial light curve PSD at each frequency, the confidence level of the QPO was estimated. The blue and red dash lines in Figure \ref{Figure3} and the right panel of  Figure \ref{Figure4} represent the confidence levels of 3$\sigma$ (99.73\%) and 4$\sigma$ (99.994\%), respectively. The confidence level of the QPO signal detected by the WWZ and LSP methods exceeds 4$\sigma$. Therefore, the 31.3-day QPO signal in S5 0716+714 is highly significant.

Through the above analysis, we get that the quasi-periodic signal in S5 0716+714 is significant. We use ARIMA and SARIMA models to fit the light curve respectively, we find that the SARIMA model has $AIC$ value much better than the best ARIMA model (Figure \ref{Figure6}b), and this fact proves that the SARIMA model with strict periodic modulation component can better model the light curve. Figure \ref{Figure6}c shows the $AIC$ values given by the SARIMA model in different seasons (s values). We give 99.994\% confidence levels based on the mean and standard deviation of the $AIC$ corresponding to the SARIMA model in different seasons. It can be seen from Figure \ref{Figure6}c that the best-fitting model of the time-varying curve for $s=30$ days has a significance of over $4\sigma$. So we have reason to believe that this QPO signal originates from the physical periodic process. In the next chapter, we will analyze and discuss the physical mechanism of QPO.
\section{CONCLUSION and DISCUSSION}

In this work, we searched for  month-like quasi-periodic oscillations (QPOs) in the Fermi Large Area Telescope Light Curve Repository (LCR) catalog of 1525 sources (all of which have variability index $>21.67$). The search results showed a possible QPO in the TeV blazar S5 0716+714. We analyzed the Fermi-LAT gamma-ray (0.1-300GeV) light curve of this source and found that a QPO of 31.3$\pm$1.8 days of this TeV blazar in MJD 55918-56137 with a FAP of less than 0.01\%. The probability that the detection due to a false alarm in a sample of 1525 sources is $1-(1-FAP)^{1525}\sim10\%$ which turns out to be roughly 10\%. In astrophysical terms, this 10\% false alarm is acceptable. The in-depth analysis found that the variations can be modeled as a $\sim$ 30-day strictly periodic component superposed on a stochastic autoregressive variability. Together these can be considered Quasi-Periodic Oscillator (QPO) behavior. The presence of both components is established independently using wavelet analysis (with the weighted wavelet Z-transform), Fourier analysis (with the Lomb-Scargle periodogram), and autoregressive analysis (comparing ARIMA and SARIMA models).
Compared with the gamma-ray QPO signal previously reported in AGN, this QPO is the first month-like QPO with 7 cycles. The QPO only occurred during the outbreak state in June 2011 (MJD 55918-56137), so this QPO was transient. Most of the QPOs that have been reported are transient signals \citep{2021ApJ...914....1Z}. We used the simulation method \citep{2013MNRAS.433..907E} to produce $10^5$ light curves to estimate the significance of the QPO, and find the confidence level of $\sim4.1\sigma$. However, this kind of month-like transient QPO perhaps originate from the central activity of the blazar or from the influence of background photons. The month-like QPO time scale is very close to the moon's sidereal period of 27.32 days, so it may be modulated by the moon's gamma-rays. But, the QPO of S5 0716+714 is transient, while the gamma-ray modulation of the moon is persistent, so the possibility of the influence of the moon can be ruled out. Secondly, the stochastic processes can produce pseudo periodic variations, but these 'phantom periodicities' typically show less than 3 cycles. When the period reaches 5 cycles, it is relatively straightforward to distinguish between potential physical cycles and stochastic processes \citep{2016MNRAS.461.3145V}. While the QPO behavior in S5 0716+714 appears in MJD 55918-56137 for up to 7 cycles, so this QPO signal is likely physical periodic process. Additionally, the SARIMA model has an $AIC$ value much better than the best ARIMA model, demonstrating that a strictly periodic component must be present in addition to any QPO-like behavior of the autoregressive component. Below, we will discuss based on the fact that the QPO originated from the center's activities.

The possible physical mechanism behind QPO in AGN has been widely discussed and many models have been proposed. One possible explanation could be a binary supermassive black hole (SMBH) system, periodicity can be induced by the secondary BH by piercing the accretion disc of the primary BH during the orbit motion \citep{2008Natur.452..851V,2010MNRAS.402.2087V,2015ApJ...813L..41A,2016AJ....151...54S}. The second, jet precession model, a secondary black hole in a non-coplanar orbit generates periodic oscillations due to tidal induced rapid precession in the inner region of the primary accretion disk \citep{2000A&A...360...57R,2018MNRAS.474L..81L}. The third, Lense-Thirring precession of accretion disks \citep{1998ApJ...492L..59S}.  The above three models are often used to explain the long term quasi periods (persistent QPOs with at least year-long periods). However, S5 0716+714 is famous for its rapid and violent intra day variability \citep{2009ApJ...690..216G,2015MNRAS.452L..11W,2017MNRAS.469.2457L,2017A&A...605A..43Y,2018AJ....155...31H,2019ApJ...880..155L,2021RAA....21..102L}, and the QPO found in this source shows a month-like oscillation, so these models can not provide appropriate explanation of this short time scale QPO. The more details of models need to be explored.

As far as we know, the gamma-ray emission in the high energy bands is attributed to a shock in the blazar jet. The variability of gamma-ray flux of the blazar can be explained in the scenario of a shockwave propagating along a helical path in the jet \citep{2008Natur.452..966M,2013ApJ...768...40L,2018A&A...619A..45M}. One possible explanation for the 31.3-day QPO is that it comes from a region with enhanced emission, moving helically within the jet \citep{2008Natur.452..851V,2015ApJ...805...91M,2021MNRAS.501...50S}. When different parts of the helical jet have different angles with the line of sight, the change of relativistic beam effect will lead to significant flux change even if there is no internal change in the emission of the jet. So, the blob in the jet moves towards us, the observation angle of helical motion essentially changes periodically, resulting in the QPO of the flux \citep{1999A&A...347...30V}. Due to the postulated helical motion of the blob, the viewing angle of the blob with respect to our line of sight ($\theta_{obs}$) changes periodically with time as \citep{2017MNRAS.465..161S,2022MNRAS.510.3641R},

\begin{equation}
cos\theta_{obs}=sin\phi sin\psi cos(2\pi t/P_{obs})+cos\phi cos\psi,
\label{eq:LebsequeIp7}
\end{equation}
where $\phi$, $\psi$, and $P_{obs}$ represent the pitch angle of the helical path, the angle of the jet axis concerning our line of sight, and the observed periodicity in the light curve respectively. The Doppler factor($\delta$) varies with the viewing angle as $\delta=1/[1-\beta cos\theta_{obs}]$, where $\Gamma = 1/(1-\beta^{2})^{1/2}$ is the bulk Lorentz factor of the blob motion with $\beta = \nu_{jet}/c$. The value of $\phi$ usually fluctuates between $0^{\circ}$-$5^{\circ}$, here we choose $\phi=2^{\circ}$, $\psi = 5^{\circ}$, and $\Gamma = 8.5$ according the typical value of blazar \citep{2017MNRAS.465..161S,2018NatCo...9.4599Z}. Due to the relativistic effect and Doppler enhancement effect, the observed period $P_{obs}$ is much smaller than the actual physical period $P_{rf}$. We can estimate the physical period $P_{rf}$ by the following formula,

\begin{equation}
P_{rf}=\frac{P_{obs}}{1-\beta cos\phi cos\psi}.
\label{eq:LebsequeIp8}
\end{equation}
Based on the known parameters, $P_{rf}$ was estimate as 7.57 years, we can also estimate the distance the blob moves in a period, $D=c\beta P_{rf}cos\phi \approx 2.09pc$. The total projection distance of 7 cycles can be estimated as $D_p=7Dsin\psi \approx 1.27pc$. The parsec scale helical jets has been detected in several blazars \citep{1995ApJ...452L..91B,1996A&A...312..727V,1998ApJ...500..810T}.
The helical structure in the jet have been supported by optical polarization observations \citep{2008Natur.452..966M}, but its origin has remained unresolved. We can assume that the jet structure is generated under the action of a helical magnetic field \citep{2004ApJ...605..656V}.

In modified helical jet model, the blob moving helically in a curved jet \citep{2021MNRAS.501...50S,2022MNRAS.510.3641R}. According to this model, the inclination angle $\psi$ of the jet's axis to the line of sight are no longer constant, but a function of time $\psi(t)$ \citep{2021MNRAS.501...50S}. In our QPO time scale, the bending of the helical jet will not have too large angle, so the impact on the value of QPO can be ignored, but the flux value of the QPO signal will slightly increase or decrease due to different inclination angle $\psi$. This provides a possible explanation for the obvious flux increase in the 7th-cycle of QPO signal.

If there are simultaneous multi-band observations, it will provide important evidence for confirming the QPO. Unfortunately, S5 0716+714 did not have available data in lower energy bands in MJD 55918-56137. The appearance of QPOs in the gamma-ray light curves of blazars is a rare event. It is even more difficult to verify it in multiple bands. So, in the future, the long-term and high temporal resolution observations of blazar is very important to identify QPOs of blazar.
\section*{acknowledgments}
We thank the anonymous referee for useful and constructive comments. This study has made use of the Fermi-LAT data, obtained from the Fermi Science Support Center, provided by NASA's Goddard Space Flight Center (GSFC). The Fermi LAT Collaboration acknowledges generous on-going support from a number of agencies and institutes that have supported both the development and the operation of the LAT as well as scientific data analysis. The work is supported by the National Natural Science Foundation of China (grants 11863007, 12063007).
\section*{Sources of data and methods}

\noindent The Fermi-LAT data used in this article are available in the LAT data server at: \url{https://fermi.gsfc.nasa.gov/ssc/data/access/}

\noindent The Fermi-LAT data analysis software is available at: \url{https://fermi.gsfc.nasa.gov/ssc/data/analysis/software/}

\noindent The Fermi Large Area Telescope Light Curve Repository (LCR): \url{https://fermi.gsfc.nasa.gov/ssc/data/access/lat/LightCurveRepository/}

\noindent The weighted wavelet Z-transform(WWZ) method: \url{https://github.com/eaydin/WWZ/}

\noindent The program REDFIT3.8e: \url{https://www.marum.de/Prof.-Dr.-michael-schulz/Michael-Schulz-Software.html/}

\noindent The light curve simulation method: \url{https://github.com/samconnolly/DELightcurveSimulation/}, \url{https://github.com/lena-lin/emmanoulopoulos}

\noindent Software for evaluating SARIMA models: \url{https://www.statsmodels.org/v0.13.0/examples/notebooks/generated/statespace_sarimax_stata.html}



\appendix
\section*{Simple presentation of the SARIMA model}
As we all know, the light curve of a blazar is a time series array, which we define as $F(t)$. First, we construct an autoregressive (AR) model for blazar's light curve,

\begin{equation}
F(t_i)=\sum^p_{j=1}\theta_{j}F(t_{i-j})+\epsilon(t_{i}),
\label{eq:LebsequeIp7}
\end{equation}
where $t_i$ is the time, $\epsilon(t_{i})$ is the fluctuation or error of the model fitting, $p$ is the time lag (the order of AR) of the past flux affecting the current flux, and $\theta_{j}$ is the coefficient of the AR model. Second, we build a moving average (MA) model for the light curve,

\begin{equation}
F(t_i)=\sum^q_{j=1}\phi_{j}\epsilon(t_{i-j})+\epsilon(t_{i}),
\label{eq:LebsequeIp7}
\end{equation}
where $q$ is the order of the MA and $\phi_{j}$ is the coefficient of the MA. Next, to consider non-stationary time series, we do d-order continuous difference (Integrated) on the light curve,

\begin{equation}
\Delta^{d}F(t_{i})=(1-L)^{d}F(t_i),
\label{eq:LebsequeIp7}
\end{equation}
where $L$ is the lag operator, $L^kF(t_{i})=F(t_{i-k})$, and $\Delta^{d}$ is the continuous difference operator, $\Delta^{d}=(1-L)^d$. Finally, AR, I, and MA are combined to form the  Autoregressive Integrated Moving Average (ARIMA) model,

\begin{equation}
\left(1-\sum\limits^p_{j=1}\theta_{j}L^{j}\right)\Delta^{d}F(t_{i})=\left(1-\sum\limits^p_{j=1}\phi_{j}L^{j}\right)\epsilon(t_{i}),
\label{eq:LebsequeIp6}
\end{equation}
where $p$, $q$, and $d$ are the order of AR, MA, and differencing respectively, lag operator defined as $L^{k}\epsilon(t_{i})=\epsilon(t_{i-k})$. For light curves with strict periodic modulation, the SARIMA model is introduced with the periodic modulation seasonal (S) component, or SARIMA$(p, d, q) \times (P, D, Q)_s$,

\begin{equation}
\left(1-\sum\limits^p_{j=1}\theta_{j}L^{j}\right)\left(1-\sum\limits^p_{j=1}\Theta_{j}L^{sj}\right)\Delta^{d}\Delta^{D}_{s}F(t_{i})=\left(1-\sum\limits^p_{j=1}\phi_{j}L^{j}\right)\left(1-\sum\limits^p_{j=1}\Phi_{j}L^{sj}\right)\epsilon(t_{i})+A(t),
\label{eq:LebsequeIp6}
\end{equation}
where $s$ is the component responsible for seasonality, $P$ and $Q$ are the orders of the seasonal AR and MA models, $\Theta_{j}$ and $\Phi_{j}$ are the coefficients of the seasonal AR and MA models, $\Delta^{D}_{s}$ is the seasonal difference operator, $\Delta^{D}_{s}F(t_{i})=(1-L^{s})^{D}F(t_i)$. For a more detailed derivation process of the ARIMA and SARIMA model, please refer to \citep{1981ApJS...45....1S,2013arXiv1302.6613A,2018FrP.....6...80F}

\end{document}